# Learnable Mixed-precision and Dimension Reduction Co-design for Low-storage Activation

Yu-Shan Tai, Cheng-Yang Chang, Chieh-Fang Teng, and An-Yeu (Andy) Wu
Graduate Institute of Electrical Engineering, National Taiwan University, Taipei, Taiwan
{clover, kevin, jeff}@access.ee.ntu.edu.tw, andywu@ntu.edu.tw

*Abstract*—Recently, deep convolutional neural networks (CNNs) have achieved many eye-catching results. However, deploying CNNs on resource-constrained edge devices is constrained by limited memory bandwidth for transmitting large intermediated data during inference, i.e., activation. Existing research utilizes mixed-precision and dimension reduction to reduce computational complexity but pays less attention to its application for activation compression. To further exploit the redundancy in activation, we propose a learnable mixed-precision and dimension reduction co-design system, which separates channels into groups and allocates specific compression policies according to their importance. In addition, the proposed dynamic searching technique enlarges search space and finds out the optimal bit-width allocation automatically. Our experimental results show that the proposed methods improve 3.54%/1.27% in accuracy and save 0.18/2.02 bits per value over existing mixed-precision methods on ResNet18 and MobileNetv2, respectively.

## I. Introduction

In recent years, deep convolutional neural networks (CNNs) have been applied in different fields and reached fabulous performance, such as face recognition [1], image classification [2], disease detection [3], and so forth. However, constrained by the limited memory storage and computation resource, it is challenging to deploy huge CNNs on edge devices directly. To alleviate this issue and extend the applicability, model compression techniques have sprung up and gained more and more attention.

Among all CNNs computation phases, data movements of activations account for the most significant part of the total energy cost. Take GoogleNet as an example, up to 68% of energy consumption comes from the data communication between deep learning accelerator (DLA) and off-chip memory [4][5]. Therefore, activation compression (AC) is emerging as an orthogonal approach to model compression. The most commonly used methods are fixed-precision quantization [6]-[10] and dimension reduction (DR) [11]-[13], which apply the same bit-width and pruning ratio to the whole model, respectively. However, both neglect the individual properties in different layers and fail to further enhance the compression ratio. Moreover, existing frameworks regard quantization and DR as independent works, without considering the interactions between these two techniques.

To further improve compressibility by exploiting layer-wise redundancy, mixed-precision (MP) quantization [15]-[19] allocates specific bit-width targeted to unique layers, as shown in Fig. 1(a). The main difficulty of mixed precision is finding the optimal bit-width combination from a large and discrete search space. To solve this issue, [18][19] employ a differentiable search architecture, relaxing search space into a continuous one and thus making gradient descent feasible.

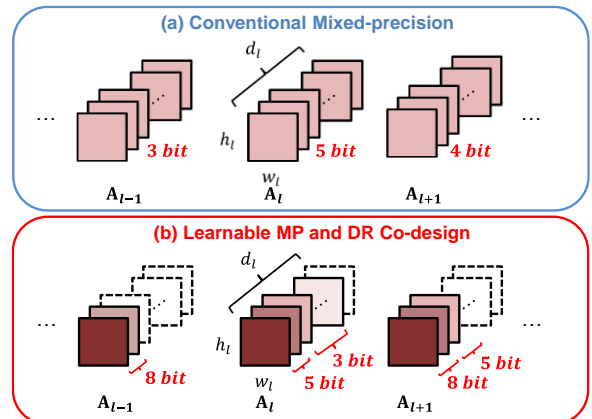

Fig. 1. Comparison of (a) conventional mixed-precision method and (b) proposed learnable mixed-precision and dimension reduction co-design system.

Although differentiable mixed-precision can take layer-wise redundancy into account and effectively enhance compressibility, there are still some issues should be addressed:

1) *Ignore channel-wise redundancy*: Since activations are generated from relatively small-sized inputs, there exists high redundancy among channels. Consequently, quantizing all channels to the same bit-width is not ideal. In an extreme case, dimension reduction should also be considered.
2) *Hand-crafted bit-width allocation*: In existing differential methods, the bit-width of each search branch is set manually. Since the optimal distribution varies with model architecture and dataset, hand-crafted design is not flexible.

In this paper, keeping the concept of differentiable search architecture, we propose a learnable mixed-precision and dimension reduction co-design system with dynamic searching, as shown in Fig. 1(b). Our main contributions are summarized as follows:

1) *First mixed-precision policy searching method for AC*: To the best of our knowledge, previous works only applied mixed-precision to reduce computation complexity. We are the first to apply this technique to minimizing activation movement.

This research work is financially supported in part by Novatek, under grant 110HT945009, and in part by Ministry of Science and Technology, Taiwan, under grant MOST 110-2218-E-002-034-MBK.

2) *Co-optimization of group-wise MP and DR*: We propose a group-wise mixed-precision system to consider channel-wise redundancy and further include DR to explore the benefits of co-optimization.
3) *Dynamic bit-width searching*: Instead of a manual setting, we propose a dynamic bit-width searching strategy to reach suitable bit-width allocation automatically. Combined with the aforementioned techniques, our simulation results get 69.5%/70.71% with average 2.56/2.62 bits per value on Resnet18/MobileNetV2.

The remainder of this paper is organized as follows: Section II briefly introduces the existing activation compression methods. Section III presents the proposed learnable dimension reduction and mixed-precision co-design system and dynamic bit-width searching technique. Section IV shows the experimental results and analysis. Finally, we conclude this paper in Section V.

## II. RELATED WORK

### A. Mixed-Precision (MP)

Due to its simple implementation and fabulous performance, quantization is popularly used for compression. Nevertheless, without considering the potential difference existing among layers, fixed-precision quantization ends up with a suboptimal trade-off between accuracy and compressibility [6]-[10]. To consider layer-wise redundancy variance and enhance compression ratio, mixed-precision gains more attraction recently [15]-[19].

However, the vast and non-differentiable search space makes it challenging to find the optimal bit-width allocation. Related research can be divided into two scopes, which are non-differentiable and differentiable ones. The former, [16][17] resorts reinforcement learning to train an agent to determine suitable policy, while [15] utilizes evolutionary algorithm to jointly search architecture, pruning ratio, and bit-width. Nevertheless, these methods require long search time. Consequently, differentiable ways [18][19], which force the model to automatically learn suitable bit-width allocation, are introduced to serve as another practical alternative for mixed-precision.

In [18], the authors design a hypernet composed of $N^\alpha$ and $N^\beta$ parallel branches for weights and activations, respectively. Each branch denotes a bit-width option for exploration, and the output during the training phase is the weighted sum of them, which can be formulated as follows:

$$y = \sum_{i=1}^{N^\alpha} \pi_i^\alpha Q_i^\alpha(\mathbf{W}) * (\sum_{j=1}^{N^\beta} \pi_j^\beta Q_j^\beta(\mathbf{A})), \quad (1)$$

$$\pi_i^\alpha = \frac{\exp(\alpha_i)}{\sum_k \exp(\alpha_k)}, \pi_j^\beta = \frac{\exp(\beta_j)}{\sum_k \exp(\beta_k)}, \quad (2)$$

where $\mathbf{W}$ and $\mathbf{A}$ denote the weight and activation, $Q_i^\alpha$ and $Q_j^\beta$ stands for uniform quantization operator with bit-width $b_i^\alpha$ and $b_i^\beta$, and $\alpha_i/\beta_i$ are the corresponding architecture parameters. Through $\alpha_i$ and $\beta_i$, the search space relaxes to a continuous version, enabling optimization by gradient descent. Moreover, since the objective of [18] is to decrease computation complexity, bit operations are added as a penalty loss to orient the training direction. During inference, only the branch with the highest $\pi_i^\alpha$ or $\pi_j^\beta$ would be assigned as the final bit-width.

Though differentiable mixed-precision successfully allocates layer-wise bit-width efficiently, there is still some room for improvement. Since activations are high dimensional vectors generated by small-sized input data, there exists high dependency among channels of the same layer. However, existing methods neglect channel-wise redundancy. Second, the search space is confined by the hand-crafted bit-width of branches, i.e., $b_i^\alpha$ and $b_i^\beta$. In fact, the optimal bit-width distribution varies with model architecture and objective dataset. Simply enlarging the number of branches, $N^\alpha$ and $N^\beta$, not only leads to extraordinary memory overhead but also suffers accuracy loss due to unconcentrated distribution of $\alpha_i/\beta_i$ and sharp bit-width drops.

### B. Transform-based Dimension Reduction (DR)

Since most compression methods are sensitive to data sparsity, they fail to compress activation well due to its high dependency on input. To overcome the restriction, transform-based methods are introduced. By applying suitable domain transformation, data can be decorrelated into important and unimportant components. After that, the latter can be removed without sacrificing model performance.

Principle component analysis (PCA) is one of the most commonly used transformations since it can customize to different input data and reach optimal decorrelation [11]-[14]. In [14], the authors propose greedy-based DR and design a selection metric, which can be specified as:

$$S_l = \Delta accuracy/\Delta N, \quad (3)$$

where $\Delta accuracy$ and $\Delta N$ represent the accuracy reduction and saved storage after operating DR on layer $l$, respectively. Small $S_l$ suggests huge memory requirement reduction while sacrificing minor accuracy. In each iteration, the least significant channel of the layer with minimal $S_l$ would be removed until reaching the desired compression ratio. The calculation of the selection metric will be elaborated in the following context.

Let $\mathbf{A}$ stand for the activation, whose size is $n \times d \times w \times h$ and $n, d, w, h$ represent the batch size, channel number, width, and height, respectively. To further fuse the transformation matrix into convolution and batch normalization by matrix folding, $\mathbf{A}$ would need to be reshaped to $\mathbf{A}_c \in d \times (n \times w \times h)$ first. Subsequently, we can obtain the corresponding transformation matrix $\mathbf{U}$ by PCA:

$$\mathbf{U}, \mathbf{\Sigma} = PCA(\mathbf{A}_c), \quad (4)$$

where $\mathbf{U}$ stands for the orthogonal basis and $\mathbf{\Sigma} = \{\sigma_1^2, \sigma_2^2, \ldots, \sigma_d^2\}$ are its corresponding eigenvalues.

After multiplying $\mathbf{A}_c$ with $\mathbf{U}$ and reshape, the transformed activation $\mathbf{A}'$ can be obtained, whose channels are sorted according to their importance. Next, quantization would be conducted before sending $\mathbf{A}'$ to off-chip memory. Consequently, the saved storage after removing one channel can be approximated as the size of a single feature map:

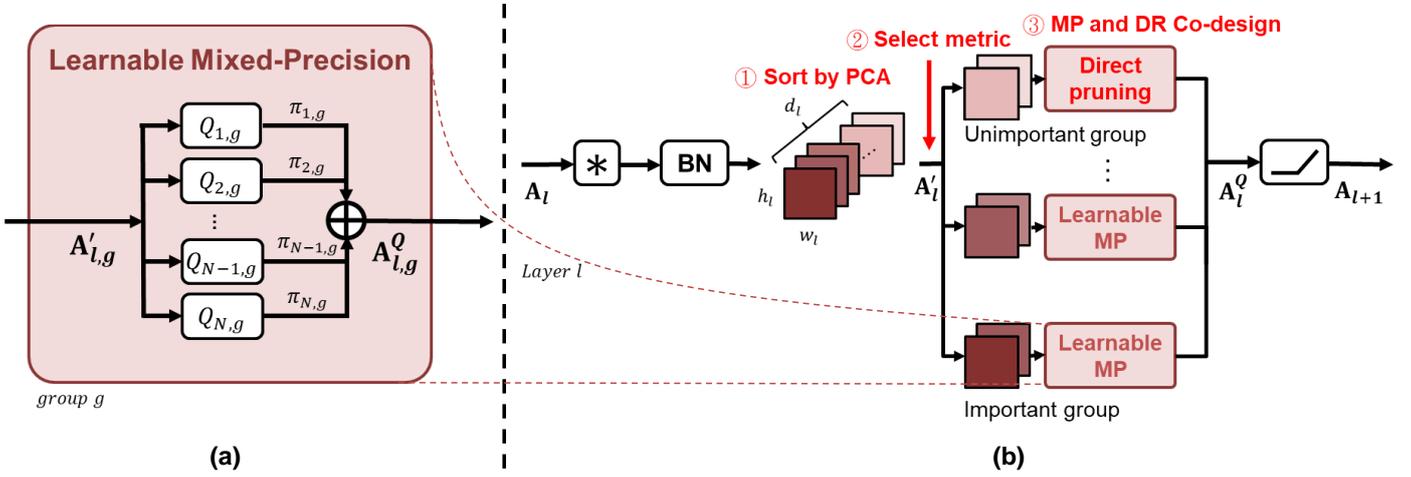

Fig. 2. (a) Learnable mixed-precision module. (b) Group-wise mixed-precision and dimension reduction co-design.

$$\Delta N = w \times h. \quad (5)$$

As for the accuracy drop estimation, [14] utilize the ratio of eigenvalue on layer $l$ to approximate the impact induced by DR:

$$\Delta accuracy \approx \sigma_{l,d_l'} / \sum_{c=1}^{d_l'} \sigma_{l,c} \quad (6)$$

were $d_l'$ stands for the reduced dimension, and $d_l'$ would update to $d_l' - 1$ if layer $l$ is selected to operate DR.

Prior results show that the selection metric is a simple yet effective strategy to achieve a better trade-off between model accuracy and compressibility. For more details, please refer to [14]. In Section III, we would further extend its functionality by enabling co-optimization of MP and DR.

## III. PROPOSED LEARNABLE MIXED-PRECISION AND DIMENSION REDUCTION CO-DESIGN SYSTEM WITH DYNAMIC BIT-WIDTH SEARCHING

In this section, we first introduce learnable mixed-precision module. Then, we elaborate on our proposed MP and DR co-design framework, which comprises group-wise independent learnable MP modules. Finally, we present dynamic bit-width searching, which is conducted on each learnable MP module.

### A. Learnable Mixed-precision Module

To achieve activation compression, we introduce a learnable MP module based on the prior differentiable method [18], which is illustrated in Fig. 2(a). Since our target is to reduce data movement of activations, we only focus on MP for activations. Therefore, we use $N$, $Q_{i,g}$, $b_{i,g}$, $\beta_{i,g}$ and $\pi_{i,g}$ to replace $N^\beta$, $Q_i^\beta$, $b_i^\beta$, $\beta_i$ and $\pi_i^\beta$ mentioned in Section II, where $\{b_{1,g}, b_{2,g}, \dots, b_{N,g}\}$ are sorted in ascending order, and script $g$ denotes group $g$ out of $G$ groups in total. The input of the learnable MP module is the transformed activation $\mathbf{A}'_{l,g}$ with size $n \times d_{l,g} \times w_l \times h_l$, and the detail of group partition would be specified later. The output $\mathbf{A}^Q_{l,g}$ during the training phase can be formulated as the weighted sum of all branches:

$$\mathbf{A}^Q_{l,g} = \sum_{i=1}^{N} \pi_{i,g} Q_{i,g}(\mathbf{A}'_{l,g}). \quad (7)$$

As for the inference phase, $\mathbf{A}^Q_{l,g}$ can be specified as the output of the branch with the largest $\pi_{i,g}$:

$$\mathbf{A}^Q_{l,g} = Q_{k,g}(\mathbf{A}'_{l,g}), \quad k = \underset{i}{\mathrm{argmax}}(\pi_{i,g}). \quad (8)$$

Unlike [18], we focus on activation compression rather than complexity reduction. Therefore, we design a memory overhead loss $\mathcal{L}_{memory}$ to encourage low bit-width allocation, which is the expected number of bits of activations over total $L$ layers:

$$\mathcal{L}_{memory} = p \sum_{l=1}^{L} \sum_{g=1}^{G} \sum_{i=1}^{N} \pi_{i,g} b_{i,g} d_{l,g} h_l w_l, \quad (9)$$

where $p$ is a hyperparameter to adjust the penalty strength. Moreover, we adopt knowledge distillation to learn from the output distribution without labeled data:

$$\mathcal{L}_{KD} = D_{KL}(\mathbf{o} || \mathbf{o}'), \quad (10)$$

where $D_{KL}(\cdot || \cdot)$ is the Kullback-Leibler divergence, and $\mathbf{o}$ as well as $\mathbf{o}'$ stand for the output vector of the original model and that of the compressed model. Finally, our loss function for training can be written as:

$$\mathcal{L} = \mathcal{L}_{memory} + \mathcal{L}_{KD}. \quad (11)$$

### B. Joint Optimization of Group-wise MP and DR

Though existing differentiable MP successfully allocates bit-width considering layer-wise difference [18], this strategy neglects redundancy among channels. To tackle this problem, we use the learnable MP modules to construct a group-wise MP system and then further co-optimize with DR to enhance compressibility.

Our framework is shown in Fig. 2(b), which contains three steps: (1) Sort channel importance by PCA, (2) Group partition by selection metric, and (3) Group-wise MP and DR co-design.

*1) Sort channel importance by PCA:* To save computation, we would use a small amount of sample data to pre-compute PCA transformation matrix $\mathbf{U}$ first. Then, after undergoing convolution ($*$) and batch normalization (BN), the transformed activation $\mathbf{A}'_l$ can be obtained after multiplied by $\mathbf{U}$.

*2) Group partition by selection metric:* With the transformed activation $\mathbf{A}'_l$, we can further divide channels into groups according to their compression priority. Instead of removing selected channels directly, we extend the functionality of the selection metric [14] to achieve group partition regarding both accuracy drop and compression strength. By iterative selection, we can identify the compression priority of channels given the order they are selected, which indicates their importance. Channels chosen in the early stage are seen as unimportant ones, while those selected in the later stage are thought to be more important. Consequently, we can separate $\mathbf{A}'_l$ into $\{\mathbf{A}'_{l,1}, \mathbf{A}'_{l,2}, ..., \mathbf{A}'_{l,G}\}$ for group 1 to G by setting different thresholds, while group 1 contains the most significant channels, and group G is comprised of the most unimportant ones. As for the thresholds for the selection metric, we first observe the limitation of the model under DR to determine the number of channels assigned to group G, and then empirically find the optimal partition.

*3) Group-wise MP and DR co-design:* After obtaining channels grouped by importance, we can apply different compression policies to exploit channel redundancy. Since members in group G are those most suitable for dimension reduction, we directly prune these channels, i.e., the values of these channels are seen as zero with no memory overhead. As for the other groups, independent learnable MP modules are operated to search group-wise optimal bit-width allocation. Finally, the output $\mathbf{A}^Q_l$ can be reformulated by concatenating $\mathbf{A}^Q_{l,g}$ from group 1 to G-1.

Since the steps would repeat each iteration during training, $\mathbf{U}$ and group partition would be updated periodically. Therefore, the channel belonging to group G might change to another group in the next iteration. Accordingly, models can dynamically modify the compression policy given current channel importance during the training phase.

*C. Dynamic Bit-width Searching*

From here, we have achieved a co-optimization system combining MP and DR together. However, the optimal bit-width allocation depends on the model architecture, target dataset, and the characteristic among groups and layers. Increasing $N$ would lead to extra computation burden and inefficient training due to unconcentrated $\pi_{i,g}$ and sharp bit-width conversion. In consequence, previous hand-crafted designs for $b_{i,g}$ are not flexible in the real case. To solve this issue, we propose a dynamic bit-width search technique, which enlarges search space without additional overheads.

Instead of manual setting for the distribution of $b_{i,g}$, we initialize each group with $N = 3$ and $\{b_{1,g}, b_{2,g}, b_{3,g}\} = \{6,7,8\}$ and then dynamically update $b_{i,g}$ according to the distribution of architecture parameters $\beta_{i,g}$. Since activations under 8 bits quantization are shown to maintain original performance, this assignment avoids a huge accuracy drop in the initial stage. After that, we can adjust $b_{i,g}$ according to the distribution of $\beta_{i,g}$. For example, as illustrated in Fig. 3, if the maximum occurred at $\beta_{1,g}$, and the minimum appeared at $\beta_{3,g}$, this implies the tendency to move toward low bit-width allocation. Therefore, we can update the search space:

$$\{b_{1,g}, b_{2,g}, b_{3,g}\} \leftarrow \{b_{1,g} - 1, b_{1,g}, b_{2,g}\}, \quad (12)$$

$$\{\beta_{1,g}, \beta_{2,g}, \beta_{3,g}\} \leftarrow \{\beta_{2,g}, \beta_{1,g}, \beta_{2,g}\}. \quad (13)$$

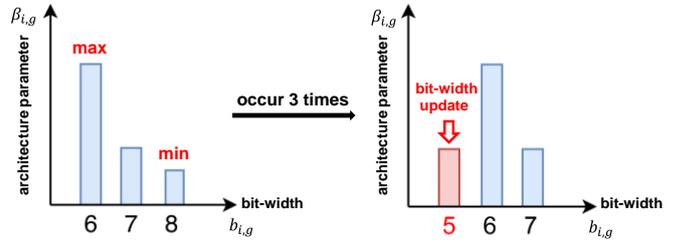

Fig. 3. Dynamic bit-width searching to enlarge search space.

We replace the branch with the largest bit-width with a small bit-width one and then update $\beta_{i,g}$ to a bell-shaped distribution. For stable training, we update branches only when the model shows a strong tendency to reduce bit-width, which can be thought as the number of occurrences of the relation $\beta_{1,g} > \beta_{2,g} > \beta_{3,g}$ reaches predefined patience, which is set as three in our following experiments.

Utilizing dynamic bit-width searching, we can enlarge the search space to various bit-width combinations with fixed $N$. Moreover, this approach forces bit-width to decrease continuously, which preserves model accuracy better.

## IV. EXPERIMENTAL RESULTS AND ANALYSIS

In the following experiments, we simulate our proposed method on pre-trained ResNet18 and MobileNetV2 with weights quantized to 8 bits. The dataset we implement is ImageNet (ILSVRC 2012) [20]. We randomly sample 50,000 out of 1,281,167 training data to fine-tune our models. For RestNet18, we set the learning rate as 1e-3 and the number of epochs is 15. As for MobileNetV2, the learning rate is 1e-4 and the number of epochs is 30. The computation is negligible compared with total model training. The batch size, penalty $p$, patience for dynamic searching, and $G$ are set as 64, [0.01~0.2], 3, and [2,3,4], respectively. We use stochastic gradient descent (SGD) as our optimizer. All the experiments are operated with PyTorch1.9.0 and Python3.9.

Since previous research [18] emphasize complexity reduction, we make some modification to prior work for a fair comparison. First, we use our proposed memory overhead loss mentioned in Section III to replace bit operations as the penalty. Next, bit-widths for exploration are set as $\{b_1, b_2, b_3, b_4\} = \{2,3,4,5\}$ for ResNet18 and $\{b_1, b_2, b_3, b_4\} = \{3,4,5,8\}$ for MobileNetV2, which are the empirical optimal allocation under our settings. Lastly, instead of training from scratch, we fine-tune a pretrained model with fewer data and batch size. After

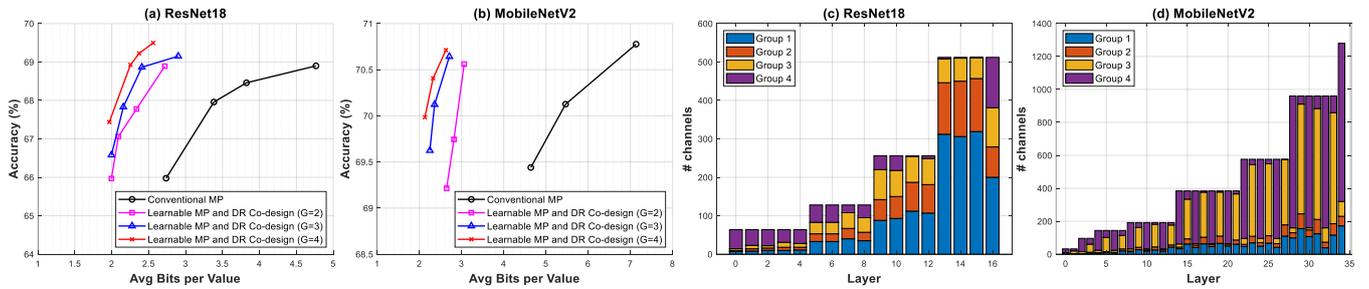
Fig. 4. Comparison of the proposed learnable MP and DR co-design system with conventional MP on (a) ResNet18 and (b) MobileNetV2, and the channel distribution of each group on (c) ResNet18 and (d) MobileNetV2 with G=4.

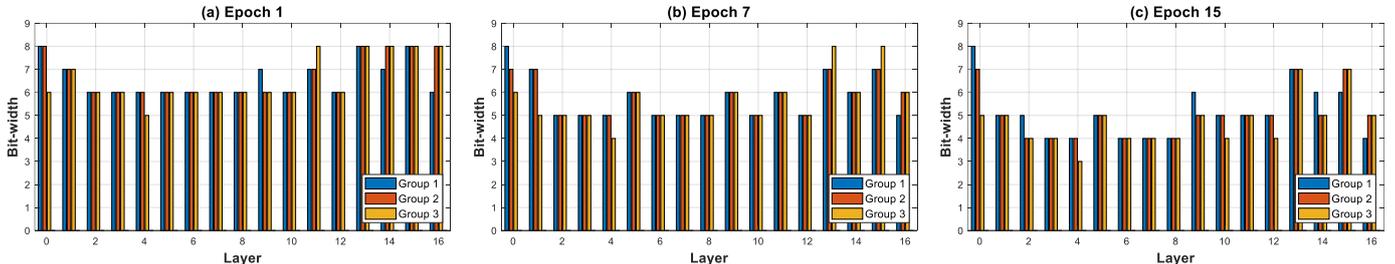
Fig. 5. Dynamic group-wise bit-width distribution of ResNet18 (G=4): (a) Epoch 1, (b) Epoch 7, and (c) Epoch 15.

that, we call the modified version conventional MP and use it as our baseline.

### A. Comparison of Proposed Learnable MP and DR Co-design System and Prior Work

In this section, we compare the compression performance of the proposed learnable MP and DR co-design system with the conventional MP approach. The simulation results are shown in Fig. 4 (a) and (b). We can observe that the accuracy of conventional MP drops hugely after the average bits per value declines to less than 3 bits and 5 bits for ResNet18 and MobileNetV2, which drops to 65.98%/69.44% with average 2.74/4.64 bits per value. However, the proposed learnable MP and DR co-design framework outperforms conventional MP a lot regardless of G. For the case of G=4, we even reach 69.5%/70.71% with average 2.56/2.62 bits per value on Resnet18/MobileNetV2. Additionally, simulation results show that a larger value of G brings a better trade-off between compression ratio and model accuracy. Since division with more groups implies smaller granularity for MP, it is more probable to capture channel-wise differences better.

### B. Visualization and Analysis of Compression Policy

In this section, we will further analyze the compression policy by visualization of the group partition and bit-width distribution. For dimension reduction, we demonstrate the number of channels of each group in Fig. 4 (c) and (d), and note that channels in group 4 would be pruned directly. For ResNet18, we can observe that pruning channels are concentrated in the shallow layers. Since the activation size in shallow layers is larger than deeper layers, they tend to get a small selection metric. As for MobileNeV2, aggressive dimension reduction occurs at depthwise layers, which implies more redundancy in depthwise layers than in pointwise layers.

As for mixed-precision, we show the group-wise bit-width distribution during the early, middle, and final training stage of ResNet18 in Fig. 5. Since group 4 is directly pruned, it is unnecessary to specify its bit-width. In epoch 1, the bit-width distribution is predefined in {6, 7, 8}. After training for more epochs, we can notice the bit-width distribution declines continuously. For the final bit-width assignments, as shown in Fig. 5(c), the distribution varies dramatically among different groups and layers, ranging from 3 bits to 8 bits. This simulation result indicates the optimal bit-width differs from layer to layer. With the proposed dynamic bit-width searching, we can enlarge the search space and automatically allocate suitable bit-width.

## V. CONCLUSIONS

In this work, we propose a learnable mixed-precision and dimension reduction co-design system with dynamic searching, which is the first work that targets mixed-precision for activation compression. By further considering dimension reduction and channel-wise differences, our method reaches a superior trade-off between model accuracy and compressibility compared with the conventional MP counterpart. Moreover, we design a dynamic bit-width searching mechanism to extend search space and automatically determine the optimal bit-width allocation. Experimental results show that our method achieves 69.5%/70.71% with average 2.56/2.62 bits per value on Resnet18/MobileNetV2.